\documentclass[aps,prl,twocolumn,showpacs,floatfix,superscriptaddress]{revtex4}
\usepackage{graphicx}
\usepackage{multirow}
\usepackage{amssymb}
\usepackage{amsmath}
\usepackage{color}
\usepackage{url}

\newcommand{\ket}[1]{\vert#1\rangle}
\newcommand{\wbar}{\overline\omega_{ge}}
\newcommand{\eul}[1]{\mathrm{e}^{#1}}
\newcommand{\eq}[1]{(\ref{#1})}

\begin{document}

\title{First-order sideband transitions with flux-driven asymmetric transmon qubits}

\author{J. D. Strand}
\affiliation{Department of Physics, Syracuse University, Syracuse, NY 13244-1130, USA}
\author{Matthew Ware}
\affiliation{Department of Physics, Syracuse University, Syracuse, NY 13244-1130, USA}
\author{F\'elix Beaudoin}
\affiliation{Department of Physics, McGill University, Montr\'eal, Qu\'ebec, Canada H3A 2T8}
\author{T. A. Ohki}
\affiliation{Raytheon BBN Technologies, Cambridge, MA 02138, USA}
\author{B. R. Johnson}
\affiliation{Raytheon BBN Technologies, Cambridge, MA 02138, USA}
\author{Alexandre Blais}
\affiliation{D\'epartement de Physique, Universit\'e de Sherbrooke, Sherbrooke, Qu\'ebec, Canada J1K 2R1}
\author{B. L. T. Plourde}
\email{bplourde@phy.syr.edu}
\affiliation{Department of Physics, Syracuse University, Syracuse, NY 13244-1130, USA}

\date{\today}


\begin{abstract}
We demonstrate rapid, first-order sideband transitions between a superconducting resonator and a frequency-modulated transmon qubit. The qubit contains a substantial asymmetry between its Josephson junctions leading to a linear portion of the energy band near the resonator frequency. The sideband transitions are driven with a magnetic flux signal of a few hundred MHz coupled to the qubit. This modulates the qubit splitting at a frequency near the detuning between the dressed qubit and resonator frequencies, leading to rates up to 85 MHz for exchanging quanta between the qubit and resonator.
\end{abstract}

\maketitle

Josephson-junction based qubits coupled to superconducting microwave resonant cavities provide an attractive route towards 
quantum information processing as well as a flexible architecture for exploring QED with electrical circuits and microwave photons ~\cite{Wallraff2004}. With multiple qubits coupled to a common resonator, it has been possible to implement 
multi-qubit gates with various schemes relying on the exchange of excitations between the qubits and resonator. 
For example, rapid magnetic flux pulses can be used to bring the qubits sequentially into resonance with the resonator for swapping real excitations \cite{Hofheinz2008, Mariantoni2011, Neeley2010}. Alternatively, multi-qubit interactions can be produced through virtual excitations between the qubits mediated by the resonator ~\cite{Majer2007, DiCarlo2010, Fedorov2011} or, in a scheme known as cross-resonance, by driving one qubit at the transition frequency of a second qubit coupled to the same resonator ~\cite{Rigetti2010, Chow2011}.
Yet another approach relies on sideband transitions first proposed for ion-trap quantum computing by Cirac and Zoller ~\cite{Cirac1995, Schmidt2003}. In the solid state, such sideband interactions have been investigated with both superconducting flux qubits \cite{Liu2007, Chiorescu2004} and Cooper-pair box qubits \cite{Wallraff2007, Blais2007} and resonators. Two-photon, or second-order, sideband transitions were used to implement a two-qubit gate between two voltage-driven superconducting transmon qubits ~\cite{Leek2009}. In this case, the second-order nature of the sideband transition limited the maximum possible gate speed. 

A recent theoretical investigation proposed an alternative approach for generating sideband transitions where the qubit transition energy is modulated near the qubit-resonator detuning frequency~\cite{Beaudoin2012}. Such a process was shown to lead to first-order sideband transitions, thus allowing for the possibility of significantly faster gates. In this Rapid Communication, we demonstrate this idea using a transmon qubit~\cite{Koch2007} and a digitally synthesized qubit frequency control (FC) waveform.  While our qubit-resonator exchange operations can be quite fast, with frequencies approaching 100 MHz, the frequency for driving the sideband can be made quite low, limited only by a reduced qubit lifetime due to the Purcell effect~\cite{Houck2008} as the qubit gets too close to the resonator.  At these low qubit-resonator detuning frequencies, typically a few hundred MHz in our experiment, expensive microwave generators are not required for the sideband drives, simplifying the control electronics and making the process more scalable.

\begin{figure}
	\includegraphics[width=0.5\textwidth]{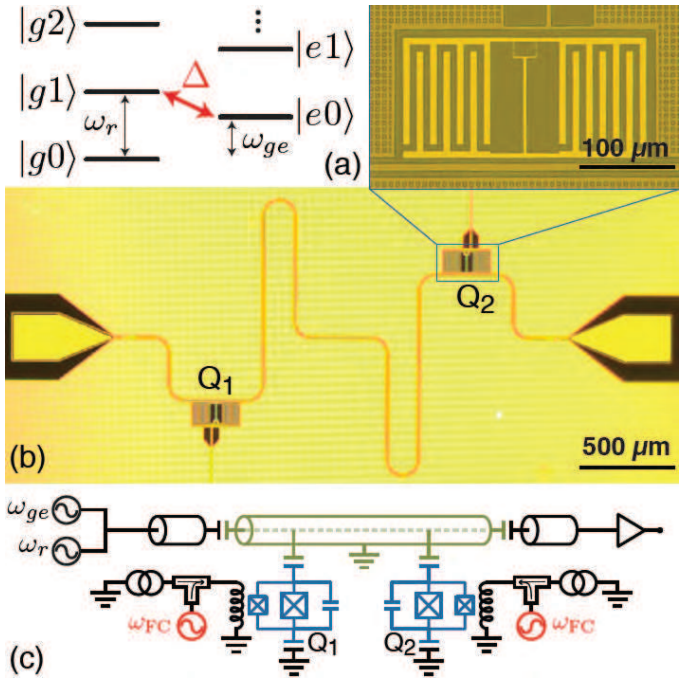}
	\caption{(color online) (a) Schematic of energy levels in a combined qubit-resonator system, showing first-order red sideband transition. (b) Optical microscope image with inset showing expanded view of one of the qubits. The terminations of the flux-bias lines for both qubits are visible, and they are used for both dc bias and FC signals. (c) Schematic of qubit-cavity layout and signal paths.}
	\label{fig:schem}
\end{figure}

To see the origin of this first-order effect more clearly, we first consider a simplified theoretical description. As a first approximation, taking into account only the first two levels \ of the transmon manifold $\left(\left\{\ket g,\ket e,\ket f... \right\}\right)$, the system can be modeled using the standard Jaynes-Cummings Hamiltonian
\begin{equation}\label{eq:HJC}
	H(t)=\omega_r a^\dagger a +\frac{\omega_{ge}(t)}{2}\sigma_z+g(a^\dagger\sigma_-+a\sigma_+),
\end{equation}
with $\omega_r$ the resonator frequency, $g$ the coupling strength ($\hbar=1$), $\sigma_z$ the usual Pauli matrix for the qubit, $\sigma_+$ ($\sigma_-$) the raising (lowering) operator for the qubit, and $a^{\left( \dagger \right)}$ the annihilation (creation) operator for the resonator. The qubit transition frequency $\omega_{ge}(t)$ is assumed to be modulated periodically using an external flux and its average value $\wbar$ is chosen such that $|\Delta| = |\wbar-\omega_r| \gg g$. In this dispersive regime, the qubit-photon interaction represented by the last term of Eq.~\eq{eq:HJC} only acts perturbatively. It can, however, be effectively turned on by periodically modulating the qubit transition frequency, using an external flux bias, such that one of the qubit's FC sidebands is on resonance with the resonator frequency $\omega_r$.

This can be made apparent by taking $\omega_{ge}(t)=\overline{\omega}_{ge}+\frac\varepsilon2\sin\omega_\mathrm{FC}t$ and moving to a frame defined by $U(t)=\exp[-i(\wbar t - \frac{\varepsilon}{2\omega_\mathrm{FC}}\cos \omega_\mathrm{FC} t)\,\sigma_z - i \omega_r a^\dagger a t ]$. In this rotating frame, the Hamiltonian reads 
\begin{align}
	&H'(t)= g a^\dagger\sigma_- J_0\left(\frac{\varepsilon}{\omega_\mathrm{FC}}\right) \eul{-i\Delta t}+\mathrm{H.c.}\label{eq:H:t}\\
		&\qquad+g a^\dagger\sigma_-\sum_{m=1}^\infty(-i)^m J_m\left(\frac{\varepsilon}{\omega_\mathrm{FC}}\right)\eul{i(m\omega_\mathrm{FC}-\Delta)t}+\mathrm{H.c.}\notag,
\end{align}
where $J_m(z)$ are Bessel functions of the first kind. For $|\Delta|\gg g$, the qubit-resonator coupling is suppressed up to an error $\sim (g/\Delta)^2$. However, when the modulation frequency $\omega_\mathrm{FC}$ is equal to $\Delta/m$, the qubit-photon interaction is reintroduced at a reduced rate $g J_m(\varepsilon/\omega_\mathrm{FC})$ corresponding to an m-photon red sideband transition. In principle, the maximum sideband Rabi frequency that can be reached in this way is $2g\max[J_1(\varepsilon/\omega_\mathrm{FC})]\simeq 1.16 g$ (for $\varepsilon/\omega_{FC} \sim 1.84$), corresponding to a large fraction of the bare frequency $2g$. In practice, one might prefer to limit the frequency modulations to have an amplitude smaller than $|\Delta|$. Even with $\varepsilon \sim \Delta/2$, we find a large rate $\sim g/2$. This is to be contrasted to sideband transitions generated by directly driving transitions of the qubit or the resonator. This results in a second-order process and therefore typically of frequency $\sim 0.01 g$~\cite{Blais2007}. For example, in Ref.~\cite{Leek2010} blue sideband transitions with a frequency of 10 MHz, or $\sim0.05g$, were generated. In addition, this FC approach to driving the sideband transitions allows complete control on the sideband Rabi frequency from the drive amplitude  $\varepsilon$. This is to be contrasted with direct resonant swaps between bare qubit-resonator states which are only controlled by the detuning $\Delta$. Moreover, working in the dressed basis, the interaction is completely turned off when $\varepsilon =0$. However, and in the same way as with the direct swap gate, one must be careful to consider the residual qubit-resonator interaction in the off-state, which will cause (possibly) unwanted frequency shifts and cross-resonance type interactions~\cite{Chow2011} in the presence of multiple qubits~\cite{Beaudoin2012}.


We chose to test this process with transmon qubits \cite{Koch2007,Schreier2008}, but rather than the conventional implementation with matching junction critical currents on either side of the qubit loop, we designed our qubits to have one junction that was several times larger than the other.  This asymmetry prevents the Josephson inductance across the qubit loop from diverging. Thus, at flux bias points of odd half-integer $\Phi_0$ ($\Phi_0 \equiv h/2e$), where a symmetric transmon would have a vanishing transition energy out of the ground state, the {\it asymmetric} transmon instead has non-zero minima in transition energy and thus doubles the number of flux-insensitive sweet spots. This also produces inflection points where the curvature of the energy bands vanishes, thus allowing for a nearly linear modulation of $\omega_{ge}$ with the qubit flux. This removes complications arising from curvature in the energy bands as discussed in Ref.~\cite{Beaudoin2012} where only symmetric transmons were considered.

\begin{figure}
	\includegraphics[width=0.48\textwidth]{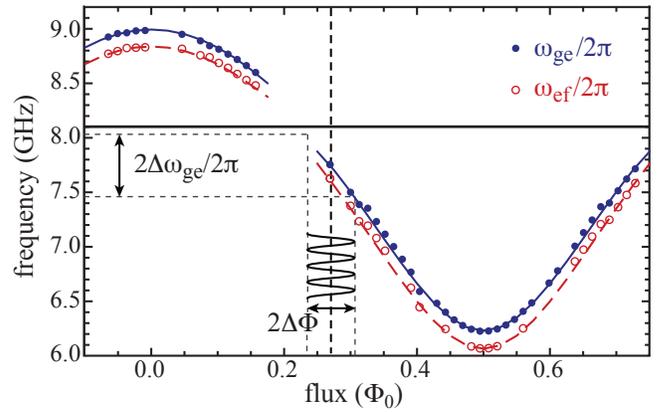}
	\caption{(color online) Spectroscopy vs. flux for Q2 showing g-e (solid blue points) and e-f (hollow red points) transition frequencies. Blue and red lines correspond to numerical fits. Heavy black line shows bare cavity resonance frequency. Vertical dashed line indicates flux bias point for sideband measurements described in subsequent figures along with ac flux drive amplitude, $2 \Delta \Phi=70.9$~m$\Phi_0$, corresponding to $2\Delta\omega_\mathrm{ge}/2\pi=572$~MHz, used in Figs. 3(c), 4(c).}
	\label{fig:spec}
\end{figure}

We used a sample consisting of two asymmetric transmon qubits capacitively coupled to the voltage antinodes of a coplanar waveguide resonator [Fig. \ref{fig:schem}(b, c)].  The cavity had a bare fundamental resonance frequency $\omega_r/2\pi=8.102\,{\rm GHz}$ and decay rate $\kappa/2\pi = 0.37\,\mathrm{MHz}$. Qubit-state measurements were performed in the high-power limit \cite{Reed2010}.  The qubits, labeled Q1 and Q2, were designed to be identical, with mutual inductances to their bias lines of $1 \,{\rm pH}$ for Q2 and $2 \,{\rm pH}$ for Q1. The qubits were excited by microwave pulses sent through the resonator, and the flux lines were used for dc flux biasing of the qubits as well as the high-speed flux modulation pulses for exciting sideband transitions. The dc flux lines included cryogenic filters before connecting to a bias-T for joining to the ac flux line, which had $20/6/10\,{\rm dB}$ of attenuation at the $4\,{\rm K}/0.7\,{\rm K}/0.03\,{\rm K}$ plates.  The distribution of cold attenuators and the flux-bias mutual inductances were chosen as a compromise to allow for a sufficient flux amplitude for high-speed modulation of the qubit energy levels with negligible Joule heating of the refrigerator while avoiding excessive dissipation coupled to the qubits from the flux-bias lines. 

We designed the qubit junction asymmetry such that $d=\left( I_{01}-I_{02} \right) / \left( I_{01}+I_{02} \right) \approx 0.5$, where $I_{01} (I_{02})$ is the critical current of the large (small) junction, to provide a frequency modulation depth of nearly $3\,{\rm GHz}$ with the upper and lower flux sweet spots on either side of the resonator frequency and the inflection point of the qubit energy bands falling slightly below $\omega_r$. This arrangement was chosen to maximize the sideband transition frequency by combining the largest flux sensitivity with a small qubit-cavity detuning, without approaching the Purcell limit. 

Figure 2 presents spectroscopy data for Q2 (solid blue and hollow red points) for the first two transitions of the transmon and the bias point  (dashed vertical line) used for our subsequent sideband measurements. The bare frequency of the resonator is indicated by the horizontal black line. The full and dashed lines are obtained by numerically diagonalizing the Rabi Hamiltonian with 4 resonator levels and 5 transmon levels. The best fits, illustrated in Fig. 2, are obtained for a Josephson energy $E_{J2}$ = 66 GHz, charging energy $E_{C2}$ = 158 MHz, junction asymmetry $d$ = 0.49, and electric-dipole coupling of the first transition $g_{ge2}$ = 129 MHz, in agreement with our measurements of the dispersive cavity shift following standard techniques \cite{Koch2007}. These qubit parameters are consistent with our design targets for the fabrication.

At the lower sweet spot, Q2 had a relaxation time $T_1=2.7\,\mu{\rm s}$ and coherence time $T_2^{*}=3.0\,\mu{\rm s}$, while at the sideband bias point these times decreased to  $T_1=1.7 \,\mu{\rm s}$, $T_2^{*}=0.6\,\mu{\rm s}$. This reduction of $ T_2^{*}$ away from the sweet spot is consistent with additional dephasing due to typical low-temperature flux-noise levels in superconducting devices and the slope of the energy band for our particular circuit \cite{Koch2007}. The coherence performance for Q1 was similar. Although we successfully performed sideband transition measurements on both Q1 and Q2, the data presented here was taken exclusively on Q2, while Q1 was kept near its lower sweet spot such that it did not factor into our results.

As illustrated in Fig.~1(a), to drive the red sideband transition, it is first necessary to prepare the qubit in its excited state. This is done with a $\pi$-pulse generated using a square pulse from a microwave generator with no additional pulse shaping. To avoid leakage, the power is chosen to be low such that a $\pi$-pulse takes $100\,{\rm ns}$. This would be too slow for eventual gate operations, but can be made faster using DRAG-type pulses~\cite{Motzoi2009}.  

\begin{figure}
		\includegraphics[width=0.5\textwidth]{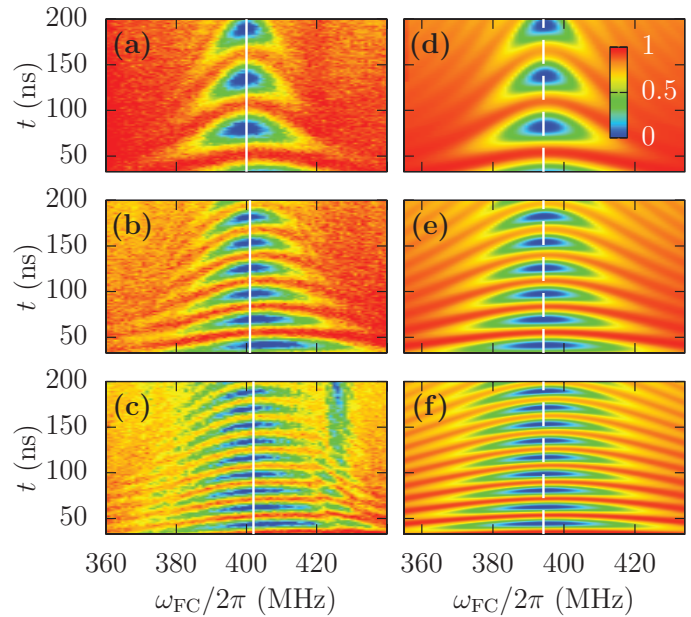}
	\caption{(color online) (a-c) Experimental data showing sideband oscillations as a function of pulse duration vs. flux-drive frequency. The amplitude of the flux pulse is reduced by (a) $10\,{\rm dB}$, (b) $4\,{\rm dB}$ relative to (c). (d-f) Corresponding numerical simulations of sideband oscillations vs. drive frequency. Vertical white lines running through each plot indicate the frequency slices used in Fig. \ref{fig:FreqVsAmpl}.}
	\label{fig:chevron}
\end{figure}

Immediately following the qubit $\pi$-pulse, a sideband pulse with a gaussian envelope and a carrier frequency $\omega_\mathrm{FC}/2\pi$ around $\Delta/2\pi = 400\,{\rm MHz}$ is sent to the flux-bias line. These sideband waveforms are generated by direct digital synthesis using a custom arbitrary pulse sequencer with no microwave electronics hardware \cite{BBN2012}. Fig.~3 shows the characteristic chevron pattern of oscillations of the qubit population with varying sideband pulse duration, which is centered on the sideband resonance. Panels (a-c) correspond to three different FC drive amplitudes, and, as expected from Eq. (2), increasing the modulation amplitude leads to faster sideband oscillations. The corresponding range of qubit frequency modulations for the largest drive amplitude is illustrated in Fig.~2. Calibration for the color scale is obtained by converting variations in the measured voltage to qubit populations using data from qubit $\pi$-pulses and $T_1$ decay. The maximum voltage achieved in a $\pi$-pulse is assumed to correspond to $P_e=1$, while the minimum value of an exponential fit to the $T_1$ decay curve sets the signal level for which $P_e=0$. These assumptions are consistent with numerical calculations and allow us to compare experimentally measured chevrons to numerical simulations.

Fig. 3(d)-(f) present simulated chevrons obtained from numerical integration of the transmon-resonator master equation taking into account damping and 5 transmon levels. Scaling of the FC drive amplitude was achieved by fitting the measurements in the low-power limit to Eq. (2) as described in the discussion of Fig. 4(d) below. All other parameters have been measured independently or obtained from the numerical fit of Fig. 2. The agreement with the measured data is excellent except for three minor differences. First, the slight asymmetry of the experimental chevrons with respect to $\omega_\mathrm{FC}$ is a result of imperfect leveling of the inherent frequency dependence of the output of the waveform synthesizer and is not present in the numerics. The feature appearing around 430 MHz in the higher power sideband oscillations appears to be due to a spurious circuit mode. It could not be reproduced from the model, even when taking into account the presence of  the far detuned Q1. Finally, the small $\sim 7\,{\rm MHz}$ shift between the center of the experimental and numerical chevrons is likely coming from the remaining imprecision in the parameter fit of Fig. 2.

\begin{figure}
	\includegraphics[width=0.5\textwidth]{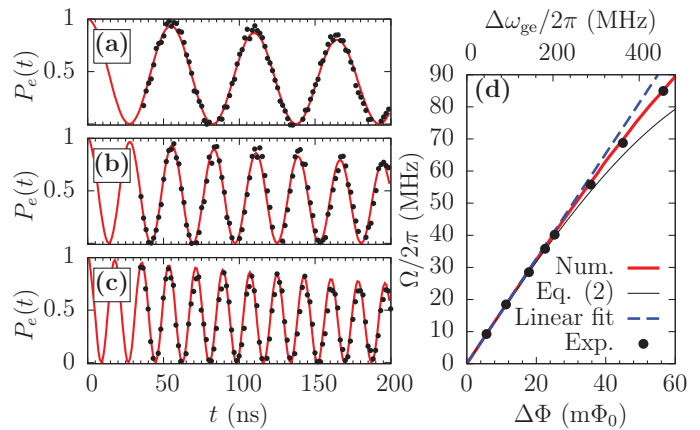}
	\caption{(color online) (a),(b),(c) Sideband oscillations corresponding to the white slices in Fig. \ref{fig:chevron}(a-c).  Experimental points correspond to black dots; numerical simulations (not fits) indicated by red lines.  (d) Sideband oscillation frequency vs. flux drive amplitude (lower horizontal axis) or corresponding frequency modulation amplitude (upper horizontal axis).  The dashed line shows a linear fit to the low frequency data points, while the red solid line indicates the theoretical dependence from the numerical simulations. The full black line shows the analytical sideband oscillation frequency from Eq.~(\ref{eq:H:t}).}
	\label{fig:FreqVsAmpl}
\end{figure}

Figure ~\ref{fig:FreqVsAmpl}(a) shows linecuts of the experimental (black dots) and numerical (full red lines) chevrons. The linecuts are taken at the frequency $\omega_\mathrm{FC}$ corresponding to the maximum-visibility sideband oscillations, indicated by the full and dashed vertical lines in Fig.~3. The agreement between the experiments and simulations is excellent. In particular, the decay rate of the oscillations can be explained by the separately measured loss of the qubit and cavity and roughly corresponds to $(\kappa+\gamma_1)/2$, where $\gamma_1$ is the bare transmon relaxation rate. This is expected for oscillations between states $|e0\rangle$ and $|g1\rangle$. It also indicates that for these powers, the visibility loss can be completely attributed to damping. The lack of experimental points at pulse widths $<30\,{\rm ns}$ is a technical limit of the present configuration of our electronics that can be improved in future experiments.Ê Ê

Figure ~\ref{fig:FreqVsAmpl}(d) shows the sideband oscillation frequency $\Omega/2\pi$ extracted from the experimental linecuts (blackÊdots) as a function of the flux-modulation amplitude $\Delta \Phi$. As expected from Eq.~\eq{eq:H:t}, whose prediction is given by the solid black line, the dependence of $\Omega$ with $\Delta \Phi$ is linear at low amplitude and deviates at larger amplitudes. Beyond this simple model with only two transmon levels, quantitative agreement is found between the measured data and numerical simulations (full red line). 
For the numerical simulations, the link between the theoretical flux modulation amplitude $\Delta \Phi$ and applied power is made by taking advantage of the linear dependence of $\Omega$ with $\Delta \Phi$ at low power. Because of this, it is possible to convert the experimental flux amplitude from arbitrary units to $\mathrm{m}\Phi_0$ using only the lowest drive amplitude for calibration.

Two effects are responsible for the deviation from linear behavior at high power: (i) the magnitude of the slope of $\omega_{ge}(\Phi)$ decreases for large excursions away from the operation point, and (ii) the relation between sideband oscillation frequency and amplitude is non-linear for strong drives, as suggested by Eq.~\eq{eq:H:t}. In the present experiments, sideband frequencies up to $85\,{\rm MHz}$ were observed without loss of visibility. In future work, with improved electronics, we should be able to access and utilize $10{\rm -ns}$ sideband $\pi$-pulses.

It is interesting to point out that there is no additional loss of visibility or overall change of behavior in both the numerical and experimental results at the largest sideband Rabi frequency although the qubit is crossing the resonator frequency in its FC excursion. The sideband Rabi frequency is thus not limited to small $\varepsilon/\Delta$. Importantly, owing to the relatively large modulation frequencies that are used, analytical and numerical calculations show that crossing another qubit or an environmental two-level system during FC excursion also would not lead to significant reduction of visibility.

We have demonstrated first-order red sideband transitions between an asymmetric transmon and a coplanar resonator, with sideband oscillation frequencies as high as $85\,{\rm MHz}$, corresponding to a full swapping of an excitation between the qubit and resonator in less than $10\,{\rm ns}$.  With these rates, following the pulse sequence described in Ref.~\cite{Beaudoin2012} for the sideband driving of two qubits coupled to a common cavity, it should be possible to perform a CNOT gate in about $30\,{\rm ns}$. This would be a major improvement in gate speed compared to previous realizations with sidebands~\cite{Leek2010} and cross-resonance~\cite{Chow2011}. These transitions were achieved with digitally-synthesized pulses, reducing the need for expensive microwave generators in this system, thus providing a favorable path towards scalability. We have also demonstrated that the asymmetric transmon, with its reduced energy-level modulation depth and natural inflection point, is an ideal component for this flux-driven sideband process. 

This research was funded by the Office of the Director of National Intelligence (ODNI), Intelligence Advanced Research Projects Activity (IARPA), through the Army Research Office under grant No. W911NF-10-1-0324. All statements of fact, opinion or conclusions contained herein are those of the authors and should not be construed as representing the official views or policies of IARPA, the ODNI, or the U.S. Government. A.B. acknowledges funding from NSERC, the Alfred P. Sloan Foundation, and CIFAR. The device fabrication involved the use of the Cornell NanoScale Facility, a member of the National Nanotechnology Infrastructure Network, which is supported by the National Science Foundation (Grant ECS-0335765).

\bibliography{Sideband_Paper}
\end{document}